\documentclass[lettersize,journal]{IEEEtran}
\usepackage{amsmath,amsfonts}
\usepackage{algorithmic}
\usepackage{algorithm}
\usepackage{array}
\usepackage[caption=false,font=normalsize,labelfont=sf,textfont=sf]{subfig}
\usepackage{textcomp}
\usepackage{stfloats}
\usepackage{url}
\usepackage{verbatim}
\usepackage{graphicx}
\usepackage{cite}
\usepackage{listings}      
\usepackage{multirow} 
\usepackage{booktabs} 
\usepackage{balance}

\begin{document}

\title{FTI-TMR: A Fault Tolerance and Isolation Algorithm for Interconnected Multicore Systems}

\author{Yiming Hu
\thanks{\textbf{Note:} This work has been submitted to the IEEE for possible publication. Copyright may be transferred without notice, after which this version may no longer be accessible.\\
version1: Draft paper \\
version2: Improved the experiment and fixed the typos.
}
}
\markboth{preprint is being prepared for submission to 
an IEEE journal}%
{ FTI-TMR A Fault Tolerance and Isolation
Algorithm for Interconnected Multicore Systems}


\maketitle

\begin{abstract}
  Two-Phase TMR conserves energy by partitioning redundancy operations into two stages and making the execution of the third task copy optional, yet it remains susceptible to permanent faults. Reactive-TMR (R-TMR) counters this by isolating faulty cores, handling both transient and permanent faults. However, the lightweight hardware required by R-TMR not only increases complexity but also becomes a single point of failure itself. To bypass isolated node constraints, this paper proposes a Fault Tolerance and Isolation TMR (FTI-TMR) algorithm for interconnected multicore systems. By constructing a stability metric to identify the most reliable nodes in the system, which then perform periodic diagnostics to isolate permanent faults. Experimental results show that FTI-TMR reduces task workload by approximately 30\% compared with baseline TMR while achieving higher permanent fault coverage.
\end{abstract}


\section{Introduction}
\IEEEPARstart{W}{ith} the continuous growth in both the number and complexity of processor cores, ensuring computational reliability in the presence of transient and permanent faults has become a fundamental requirement for hard real-time and safety-critical applications\cite{RN57, RN49}. Triple Modular Redundancy (TMR) addresses this need by executing three identical copies of a task and using majority voting on their outputs to determine the system’s final result.

However, TMR inherently incurs roughly three times the energy consumption of a non-redundant system. Recent research has increasingly focused on improving fault tolerance while minimizing energy overhead. It has also explored the trade-offs between these two objectives\cite{RN54}. Various TMR variants have been proposed to achieve this balance\cite{RN24, RN2}. For instance,\cite{RN24} introduces a Two-Phase TMR scheme in which the third copy can be skipped if the outputs of the first two copies match. Building on this concept,\cite{RN2} presents an energy-efficient Reactive TMR (R-TMR) approach that detects and isolates cores with permanent faults using lightweight hardware mechanisms and optimized scheduling policies.

Despite these advancements, TMR and its variants are largely restricted to the ``isolated node'' paradigm, relying on redundant resources within a single computing device. This paradigm presents inherent limitations: the added hardware (e.g., detection logic in R-TMR) increases design complexity and introduces new single points of failure. Moreover, if a node suffers multiple permanent core failures or significant resource degradation, its fault tolerance capability deteriorates rapidly. In essence, the reliability ceiling of these methods is bound by the hardware resources of the individual node.

The thriving of the interconnected systems paradigm provides a new opportunity to overcome these limitations. According to a Cisco report\cite{RN77}, global machine-to-machine (M2M) connections were projected to reach 14.7 billion by 2023, accounting for 50\% of all connected devices. A significant portion of these interconnected systems relies on multicore processors and graphics processing units (GPUs) to meet the ever-growing demands for performance and energy efficiency\cite{RN78}. %
However, as transistor dimensions continue to shrink and device densities increase, these components have become increasingly susceptible to aging-related degradation mechanisms, including bias temperature instability (BTI), hot-carrier injection (HCI), and electromigration\cite{RN79}. The cumulative impact of these phenomena can resulting in permanent hardware faults.

Motivated by this, this paper proposes a dynamic, adaptive fault-tolerant algorithm for interconnected systems: Fault Tolerant and Isolation TMR (FTI-TMR). The key idea is to shift fault detection from reliance on dedicated hardware to a dynamically allocatable “service” distributed across nodes. Based on a stability metric, the two most stable nodes are selected to periodically perform fault detection. This metric-driven, lightweight mechanism eliminates the need for additional hardware, avoids single points of failure, and can adaptively manage scenarios involving multiple node failures, thereby significantly enhancing overall system robustness.

Experimental results show that, compared with conventional TMR techniques, FTI-TMR reduces task workload by approximately 30\%. Compared to the state-of-the-art R-TMR\cite{RN2}, FTI-TMR achieving higher permanent fault coverage. The main contributions of this work are as follows:
\begin{itemize}
  \item{Developed and open-sourced a TMR fault tolerance simulation tool based on the Standard Task Graph (STG) dataset, simplifying the data collection and comparative experimentation process.}
  \item{We extended the experiments on Reactive TMR and demonstrated its fault-tolerance limit.}
  \item{Proposed a stability metric suitable for multicore device under TMR.}
  \item{We designed a dynamic, fault-tolerant algorithm for interconnected systems capable of tolerating transient faults and isolating permanent faults.}
\end{itemize}


\begin{figure}[!t]
  \centering
  \includegraphics[width=3.4in]{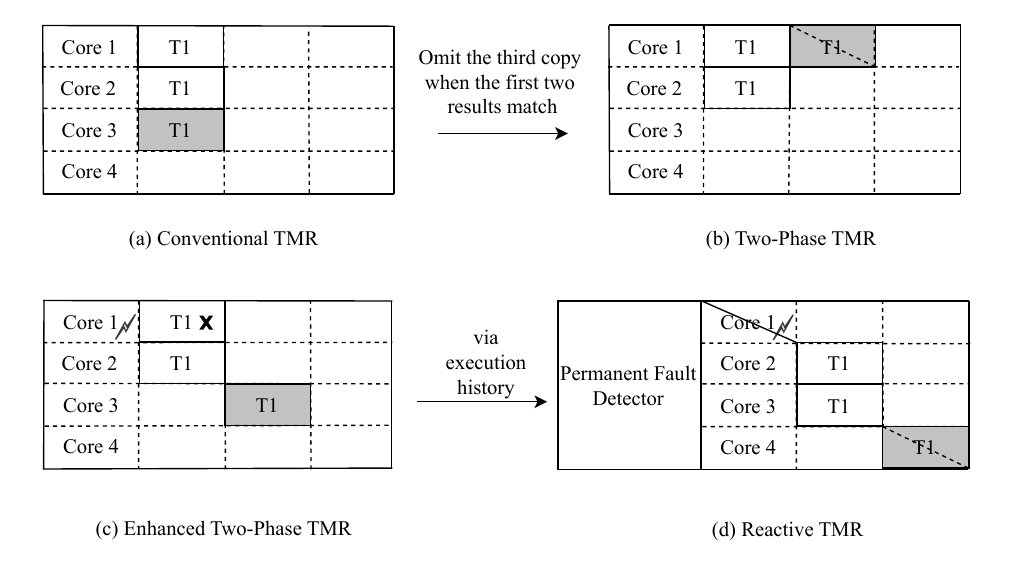}
  \caption{TMR schemes: (a) Conventional TMR:~three task copies must be executed.
  (b) Two-Phase TMR:~the third copy can be omitted under normal conditions.
  (c) Enhanced Two-Phase TMR:~task copies are distributed across different cores, allowing tolerance to permanent core faults.
  (d) Reactive TMR:~based on Enhanced Two-Phase TMR, a permanent-fault detector is introduced to disable the use of faulty cores according to execution history.}\label{fig:background}
\end{figure}
\newpage
\section{Background and \\ Related Work}\label{sec:relatedwork}
In real-time systems, fault tolerance mechanisms are among the key technologies for ensuring system reliability. Redundant execution is a widely adopted fault tolerance strategy, the core idea of which is to introduce additional computational resources to improve the system's error tolerance. N-Modular Redundancy (NMR) achieves fault tolerance by executing multiple copies of a task in parallel and using a majority vote to produce the final output. By masking errors in individual copies, this approach helps maintain correct system operation. NMR can be implemented via hardware redundancy (e.g., replicating multiple circuit modules) or software means (e.g., running multiple program instances).

Triple Modular Redundancy (TMR), as the most representative form of the NMR scheme, can effectively tolerate single-point failures through the execution and voting mechanism of three redundant copies. TMR is widely used in high-reliability scenarios such as spacecraft and nuclear power plant control systems. Under the majority voting mechanism, a TMR system requires at least two correct copies to maintain system functionality. Numerous studies\cite{RN24, RN2, RN58} have highlighted TMR’s effectiveness in improving the reliability of hard real-time systems. The TMR and its variants always focused on an isolated node paradigm, where all redundant resources are contained within a single computing device.

Recently, the thriving of interconnected systems has become a defining characteristic of modern computing infrastructures, from large-scale data centers and cloud platforms to embedded and edge devices, interconnected computing nodes now support unprecedented parallelism and resource sharing. 

Despite the existence of several studies proposing fault detection algorithms for interconnected systems, most of them remain theoretical. Their primary objective is to demonstrate, under specific system models and detection schemes, how many faulty nodes can be detected given a certain number of fault-free nodes\cite{RN74, RN75, RN76, RN80}. 

This section provides a short overview of TMR-related algorithms in multicore environments, covering conventional TMR (Section~\ref{subsec:C-TMR}), Two-Phase TMR (Section~\ref{subsec:TP-TMR}), Enhanced Two-Phase TMR (Section~\ref{subsec:TP-TMR-E}), and Reactive TMR (Section~\ref{subsec:R-TMR}). The first three approaches are implemented entirely in software, whereas Reactive TMR requires extra hardware to detect and isolate permanent faults. In addition, Section~\ref{subsec:Raft} introduces the Raft consensus algorithm from distributed systems, which serves as a key source of inspiration for the core concepts developed in this paper.

\subsection{Conventional TMR}\label{subsec:C-TMR}

Conventional TMR is the most fundamental implementation. Its execution process is illustrated in Fig.~\ref{fig:background} (a). The system executes three copies of a task and determines the final output by comparing their results. If one copy's result diverges from the other two, that copy is identified as faulty, and its output is discarded. System failure occurs only if all three copies produce mutually inconsistent results. However, in practice, the probability of concurrent failures in two or more copies (in the absence of permanent faults) is exceedingly low\cite{RN66}. Thus, conventional TMR effectively maintains system functionality in most scenarios.

\subsection{Two-Phase TMR}\label{subsec:TP-TMR}

Conventional TMR executes triples the workload of a non-redundant baseline system, introducing significant energy overhead. Because transient faults are relatively rare in practice, Two-Phase TMR divides the traditional TMR execution workflow into two phases: a mandatory phase and an on-demand phase\cite{RN24}. Two copies of each task are scheduled for execution in the mandatory phase, while the third copy, reserved for the on-demand phase, is executed only if needed.

During the mandatory phase, the system executes the two task copies and compares their results. If the outputs agree, the third task copy is skipped, thereby conserving energy. Since transient faults are uncommon\cite{RN81}, the results of the two mandatory copies typically match. This allows Two-Phase TMR to omit the on-demand phase in most execution cycles, leading to substantial overall energy savings. Fig.~\ref{fig:background} (b) illustrates an example task execution under Two-Phase TMR.

\subsection{Enhanced Two-Phase TMR}\label{subsec:TP-TMR-E}

Two-Phase TMR do not specify how task copies are allocated across processor cores. Two task copies could be executed serially on a single core. If executed serially on one core, the system's fault tolerance is compromised if that specific core experiences a permanent fault\cite{RN2}. As shown in Fig.~\ref{fig:background} (c), Enhanced Two-Phase TMR addresses this by distributing the three task copies across three distinct cores for execution, thereby tolerating permanent faults affecting any single core.

\subsection{Reactive TMR}\label{subsec:R-TMR}

Although Enhanced Two-Phase TMR tolerates single-core permanent faults, it increases the likelihood that the third task copy cannot be omitted. Building upon Enhanced Two-Phase TMR, Reactive TMR (R-TMR), illustrated in Fig.~\ref{fig:background} (d), introduces additional lightweight hardware for permanent fault detection. When it detects a core consistently producing erroneous outputs over multiple voting rounds, R-TMR classifies that core as permanently faulty core. Subsequently, a task migration mechanism reassigns tasks from the failed core to healthy ones, maintaining system correctness and availability.

Although prior work primarily demonstrated R-TMR's capability for detecting and isolating a single faulty core\cite{RN2}, our experiments in Section~\ref{subsec:ra} show that, provided the permanent fault detection hardware remains functional, R-TMR can detect and isolate any number of permanent faulty cores within the system as long as at least two fault-free cores are available. 

Despite its advantages, R-TMR has two limitations. First, the algorithm fails rapidly when fewer than two fault-free cores are available. Second, its reliance on additional hardware fault detection units (including fault history arrays, flag bits, backup scheduling policy storage, etc.) introduces a potential single point of failure; malfunction of this unit could potentially induce more severe system failures. Section\ref{subsec:Motivation} discusses the limitations of R-TMR in more detail.

\subsection{Raft}\label{subsec:Raft}

Raft is a consensus algorithm designed for managing replicated logs in distributed systems, with a primary design goal of being easy to understand\cite{RN27}. Unlike conventional algorithms like Paxos\cite{RN71}, known for their complexity and difficulty in practical implementation, Raft significantly reduces engineering complexity and cognitive load by decomposing the consensus problem into three relatively independent subproblems: leader election, log replication, and safety.

\begin{figure}[!t]
  \centering
  \includegraphics[width=3.2in]{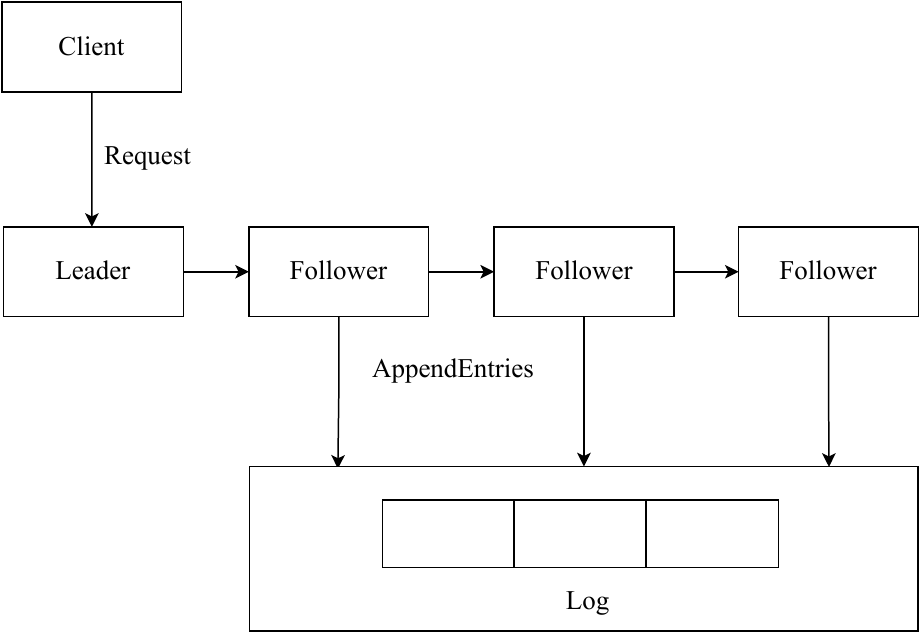}
  \caption{Raft consensus algorithm.}\label{fig:raft}
\end{figure}

In the Raft architecture, any server node is in one of three states at any given time: leader, follower, or candidate. Its operation centers around a strong leader model: for a given term, one and only one leader is elected. This leader is responsible for receiving all client requests and managing the log replication process as shown in Fig.~\ref{fig:raft}. This centralized management simplifies the algorithm's logic, ensuring log entries are appended unidirectionally from the leader to followers, avoiding complex conflict resolution mechanisms.

Raft not only provides an efficient and reliable consensus solution but has also become the de facto standard in industry for building distributed systems\cite{etcd, consul}, due to its clear modular design and strong leader model. The success of Raft and TCP/IP, compared to more complex alternatives such as Paxos or the ISO protocol stack, suggests that simplicity and understandability can sometimes outweigh theoretical sophistication, making protocols more practical for real-world deployment. The fault-tolerant mechanism proposed in this paper is deeply inspired by Raft's approaches to strong leader model and state consistency.

\section{The proposal}
This section presents a fault-tolerant architecture for interconnected multicore systems, capable of addressing both transient and permanent faults. The core idea of our approach is to assess the relative reliability of processing units based on their historical and runtime behavior, and to select two highly stable computation nodes to perform fault detection on the remaining nodes and then, to notify them to isolate permanent fault cores. We begin this section with an example: Section~\ref{subsec:Motivation} Motivation highlights a specific scenario that the recently proposed Reactive TMR (R-TMR) fails to handle. We then describe our method in detail, covering Section~\ref{subsec:sysmodel} the System Model,~\ref{subsec:htf} Handling Transient Faults,~\ref{subsec:Sm} Stability Metrics,~\ref{subsec:pdipf} Periodic Permanent faults Detection and Isolation, and Section~\ref{subsec:tgftitmr} The Guarantees of FTI-TMR.

\subsection{Motivation}\label{subsec:Motivation}
\begin{figure}[!t]
  \centering
  \includegraphics[width=3.4in]{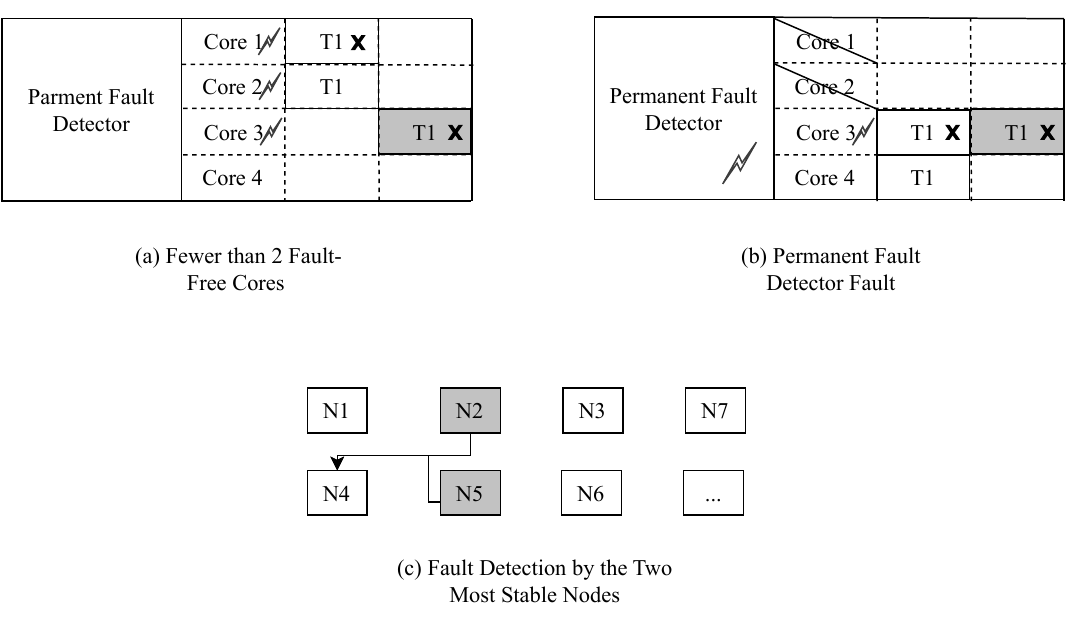}
  \caption{Limitations of R-TMR and Improvement Concept:~(a) the system fails when there is fewer than 2 fault-free cores, or (b) when the additional fault-detection hardware becomes defective. Improvement Concept: (c) the two most stable nodes are selected to perform permanent fault detection on the other nodes.}\label{fig:motivation}
\end{figure}

Fig.~\ref{fig:motivation} illustrates the execution flow of a hypothetical task T1 under R-TMR. When nodes Core1, Core2, and Core3 experience permanent faults, Fig.~\ref{fig:motivation} (a) shows that the R-TMR system fails if fewer than two fault-free cores remain. Because Core1 and Core2 are faulty, their results after T1’s execution are inconsistent with Core3, prompting the system to switch to on-demand execution of the third copy. However, since Core3 is also faulty, executing T1 on it still produces inconsistent results. The system completely fails and cannot identify the faulty cores.

Fig.~\ref{fig:motivation} (b) shows an even more severe situation: if the permanent fault detection unit is broken, the healthy cores Core1 and Core2 may be unexpectedly shut down, while the faulty Core3 remains active. Then the system cannot assign the third task copy to a different core. If two copies of T1 are executed on the faulty cores, correct voting cannot be achieved. In such a scenario, the entire algorithm inevitably fails.

In light of this, we propose an improved approach that does not rely on additional hardware units for fault detection and isolation. As illustrated in Fig.~\ref{fig:motivation} (c), within an interconnected system, the system periodically votes to select two more stable primary and secondary leader nodes, which perform fault detection on the other nodes and instruct the healthy cores to isolate the cores with permanent fault.

\subsection{The System Model}\label{subsec:sysmodel}

We consider an interconnected system comprising multi-core nodes $N_1$, $N_2$, $N_3$\dots$N_n$. The system can be either homogeneous or heterogeneous. Each node is assumed to possess at least one CPU, and each CPU contains a minimum of three cores. During normal operation, each node can independently run different applications.

A process is bound to each core on every node. These processes enable inter-core communication within a node and inter-node communication across the system. $Core_{ij}$ denotes the $j$-th core of node $N_i$. For instance, communication between $Core_{12}$ and $Core_{23}$ signifies communication between the process on the second core of node $N_1$ and the process on the third core of node $N_2$. We assume inter-node communication is based on the TCP protocol, employing encrypted channels that are secure, fault-free, and reliable. The disabling of cores within a node is formally defined under Item~\ref{item:logicalIsolation} in Section~\ref{subsec:tgftitmr}.

Both transient faults (e.g., induced by radiation or noise) and permanent faults (e.g., caused by aging or physical damage) may occur unpredictably on any core. The FTI-TMR algorithm proposed in this paper primarily targets both transient and permanent faults within interconnected multi-core systems. We assume cores with faults in the system are honest-but-faulty. This means a faulty core might exhibit non-malicious failure behaviors such as crashing, timing out, sending messages with outdated states, or producing incorrect computational results. It will not actively collude with other cores to send malicious or contradictory messages intended to deceive the system (which is known as the Byzantine generals problem\cite{RN73}). We define this as an Accidental-Behavior Fault Tolerance (ABFT) model. The reliable operation of the system relies on the fundamental premise that, for n nodes in the system, $\lfloor\frac{n}{2}\rfloor + 1$ nodes remain fault-free.

\subsection{Handling transient faults}\label{subsec:htf}
To handle transient faults during the operation of a node $N_i$, we employ Enhanced Two-Phase TMR (TP-TMR+, see Section \ref{subsec:TP-TMR-E}). This approach distributes task copies across distinct cores for execution and voting. For directed acyclic graph (DAG)-based tasks, the List Scheduling\cite{RN68} with the Longest Task First (LTF)\cite{RN69} scheduling strategy is adopted, while for independent discrete tasks, the First-Come-First-Served (FCFS) policy is applied.

disputed task is defined as one that triggers the on-demand phase due to inconsistent outputs during majority voting. When such a task is identified, the core $Core_{ij}$ increments its dispute counter $D_i$ and stores the task along with its input data in a local disputed task list, denoted as $DTList_i$. It should be noted that in a node with four cores, for instance, each core maintains its own copy of the disputed task data---resulting in four separate yet identical records across the node. The structure of $DTList_i$ is detailed in Table~\ref{table:dtlist}, where $DTList_i[0]$ corresponds to the disputed tasks recorded on the first core.
\begin{table}
  \begin{center}
  \caption{The data structure of disputed task list $DTList$}\label{table:dtlist}
  \begin{lstlisting}[frame=single,basicstyle=\ttfamily,language=Java]
    DTList = [
      [{
        task, // Disputed task
        input, // task input data on Ni
        passed, // passed test or not
        failedCount // used to exponential backoff
      }, ...], // Core 1
      [...], // Core 2
      [...], // Core 3
      ...
    ]
    \end{lstlisting}
  \end{center}
\end{table}

Both the dispute counter $D_i$ and the dispute list $DList_i$ are utilized in the computation of the node stability metric and support the permanent fault isolation.

In scenarios where a node has degraded to fewer than three active cores---making it impossible to allocate each task copy to a distinct core---the system continues to distribute task copies across the remaining cores as evenly as possible. For example, with three tasks and only two functioning cores, two tasks will inevitably be executed on the same core.

\subsection{Stability metrics}\label{subsec:Sm}
To identify nodes with high long-term operational stability, we propose a stability scoring mechanism. For each node $N_i$, the total number of tasks executed is denoted as $S_i$, and the number of disputed tasks generated is denoted as $D_i$. Let $f_i$ and $F_i$ represent the actual value and the maximum value, respectively, of other factors influencing stability (such as the value of an exponential distribution function and its maximum value of 1). This mechanism is based on the Stability Score ($SS$) computed for each node:

\begin{equation}
  \label{sseq}
  SS_i = \frac{S_i-D_i+f_i}{S_i+F_i}.
\end{equation}
Equation~\eqref{sseq} quantifies the operational stability of device $N_i$. The stability score $SS_i \in [0, 1]$ is defined such that a higher value indicates greater stability. Each core records relevant parameters ($D_i$, $f_i$, etc.) for this computation. The underlying principle of this metric is fairly simple: nodes prone to permanent faults are more likely to generate disputed tasks.

\begin{figure}[!t]
  \centering
  \includegraphics[width=3.3in]{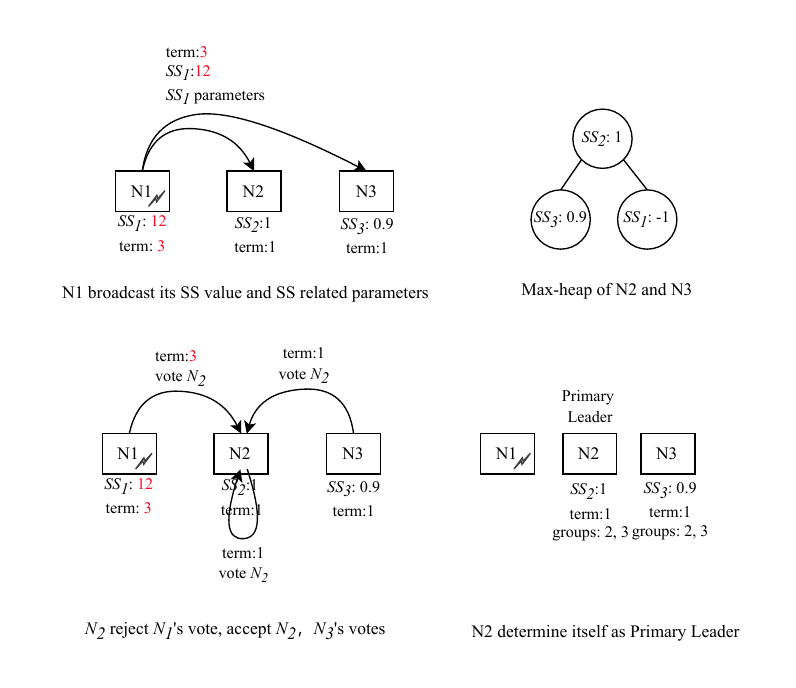}
  \caption{Primary leader election.}\label{fig:primaryLeaderElection}
\end{figure}

\begin{table}
  \begin{center}
  \caption{Strategies for addressing term Inconsistency.}\label{table:termInconsistency}
  \begin{tabular}{ c  c }
  \toprule
  Situation & Strategy \\
  \midrule
  Received & Insert $SS_i=-1$ \\
  $SS_i$ value & into max-heap \\
  \midrule
  $N_i$ under & Perform fault isolation on the $N_i$  \\
  fault isolation & if data is intact, else reject request \\ 
  \midrule
  Heartbeat of $N_v$ and $N_p$ & see Section~\ref{subsec:tgftitmr} items~\ref{item:logicalIsolation} \\
  \bottomrule 
  \end{tabular}
  \end{center}
\end{table}

\subsection{Periodic detection and isolation of permanent faults}\label{subsec:pdipf}
The periodic detection and isolation of permanent faults is initiated by each node at with a period $t$, where $t$ may be a fixed or variable duration. This process comprises two main components: Primary and secondary leader election and permanent fault detection and isolation.

\subsubsection{Primary and Secondary Leader Election}\label{subsubsec:leaderElection}
According to Equation~(1), for a node $N_i$ within the system, each local core records the round number, \textit{term}, during the permanent fault isolation phase. The \textit{term} starts from $0$ and increments by $1$ each time a voting process begins. %
All cores on all nodes attach the current \textit{term} to their requests, and the receiving nodes verify whether the received \textit{term} matches their own. This \textit{term} mechanism is a key factor in ensuring the reliability of the algorithm under the  Accidental-Behavior Fault Tolerance (ABFT) model: it guarantees that all correct cores operate within the same logical time cycle during leader election and state synchronization, effectively rejecting interference from cores whose states may be inconsistent due to crashes or restarts, thereby preserving the temporal consistency of leader election. The handling of \textit{term} inconsistencies is summarized in Table~\ref{table:termInconsistency}.

\begin{figure}[!t]
  \centering
  \includegraphics[width=3.4in]{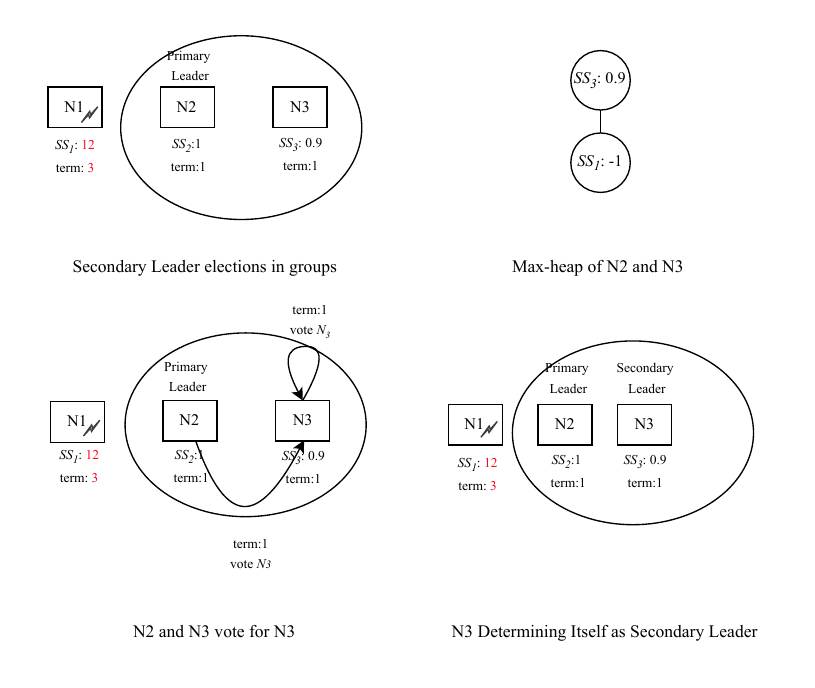}
  \caption{Secondary leader election.}\label{fig:secondaryLeaderElection}
\end{figure}

Fig.~\ref{fig:primaryLeaderElection} and~\ref{fig:secondaryLeaderElection} illustrates the leader election process. Each node introduces a random delay of 150–300~ms, as in previous work~\cite{RN27}, before a core is selected to respond and communicate with other nodes. This core is referred to as the \textit{Contact Core}. Upon computing $SS_i$, the \textit{Contact Core} broadcasts $SS_i$ along with the dependent parameters (e.g., $S_i$ and $D_i$) to the \textit{Contact Core} of other nodes. %
When node $N_j$ receives $SS_i$ and the associated parameters from node $N_i$, it executes the following logic: if the current \textit{term} matches and the verification confirms that $SS_i$ strictly corresponds to the computed result based on its parameters and lies within the range $[0,1]$, then the node and its $SS_i$ value are inserted into a max-heap according to the value $SS_i + i$ (the addition of $i$ prevents multiple nodes with identical $SS_i$ values from being inserted in the same order, which could otherwise lead to voting confusion). If the \textit{term} is inconsistent, or $SS_i$ does not match the computed result, or both conditions occur, the node is inserted into the max-heap with a value of $-1$, and subsequently $N_j$ will reject votes from nodes with $SS_i=-1$. All subsequent primary and secondary leader elections rely on the \textit{Contact Core} to perform inter-node communication and coordination.

After node $N_i$ in the system receives the $SS$ values from all nodes, it initiates the primary leader election. Node $N_i$ extracts the top element from the max-heap as the intended candidate (Primary OP) and votes for the node $N_p$ with the highest $SS$ value to be elected as the primary leader. Since the system operates correctly only when more than half of the nodes function properly, the majority of nodes will vote for $N_p$, electing it as the primary leader, as shown in Fig.~\ref{fig:primaryLeaderElection}.

Once $N_p$ receives votes from more than half of the nodes, it broadcasts a notification to all nodes in the system announcing itself as the primary leader. Upon receiving acknowledgment (Ack) messages from other nodes, $N_p$ adds them to its Group list. After receiving Acks from the majority of nodes, $N_p$ confirms itself as the primary leader. When the primary leader election phase ends, the primary leader instructs all nodes in its Group to perform a secondary leader election.

The primary leader and the nodes in the Group vote for a secondary leader $N_v$ in the same manner as in the primary leader election. $N_p$ is responsible for making judgments and submitting orders, while $N_v$ supervises and assists $N_p$'s actions, Nodes that fail to establish a primary leader before the timeout will reinitiate the primary leader election, while nodes that have already confirmed a primary leader will reject such election requests. The process of secondary leader election is illustrated in Fig.~\ref{fig:secondaryLeaderElection}.

The random selection of a \textit{Contact Core} is designed to prevent a node from inadvertently choosing a faulty core to broadcast the $SS$ value. If a node selects a faulty core as its \textit{Contact Core}, the system must handle the following three cases:

\begin{enumerate}
  \item{The faulty \textit{Contact Core} broadcasts an incorrect $SS$ value. According to equation (1), the $SS$ value should fall within the range $[0, 1]$. If the received $SS$ value is outside this range, and verification based on equation (1) fails, the receiver will assign the $SS$ value of that node to $-1$ and insert it into the max-heap. The same handling applies when the term values are inconsistent.}
  \item{The faulty \textit{Contact Core} fails to broadcast the $SS$ value. If the $SS$ value is never broadcast, the system will immediately initiate the primary leader election once the $SS$ broadcasting phase times out. Since more than half of the nodes are functioning correctly, the election can still proceed normally.}
  \item{The faulty \textit{Contact Core} falsely claims leadership. Each node maintains its intended candidate; if a node receives a leadership claim from a node different from its own intended candidate, it will reject the claim.}
\end{enumerate}

As shown in Fig.~\ref{fig:motivation} (c), the two most reliable nodes, $N_2$ and $N_5$, are designated as the primary and secondary nodes to perform fault detection for other nodes within the system, $N_2$ is making judgments and submitting orders, while $N_5$ supervises and assists $N_2$'s actions. After the timeout of the primary-secondary leader election phase, nodes that have successfully established the primary and secondary leaders subsequently enter the fault detection and isolation phase. The fault-tolerance guarantees of the FTI-TMR algorithm will be discussed in detail in Section~\ref{subsec:tgftitmr}.

\subsubsection{Permanent fault detection and isolation}\label{subsubsec:faultIsolation} 
\begin{figure*}[!t]
  \centering
  \includegraphics[width=6.8in]{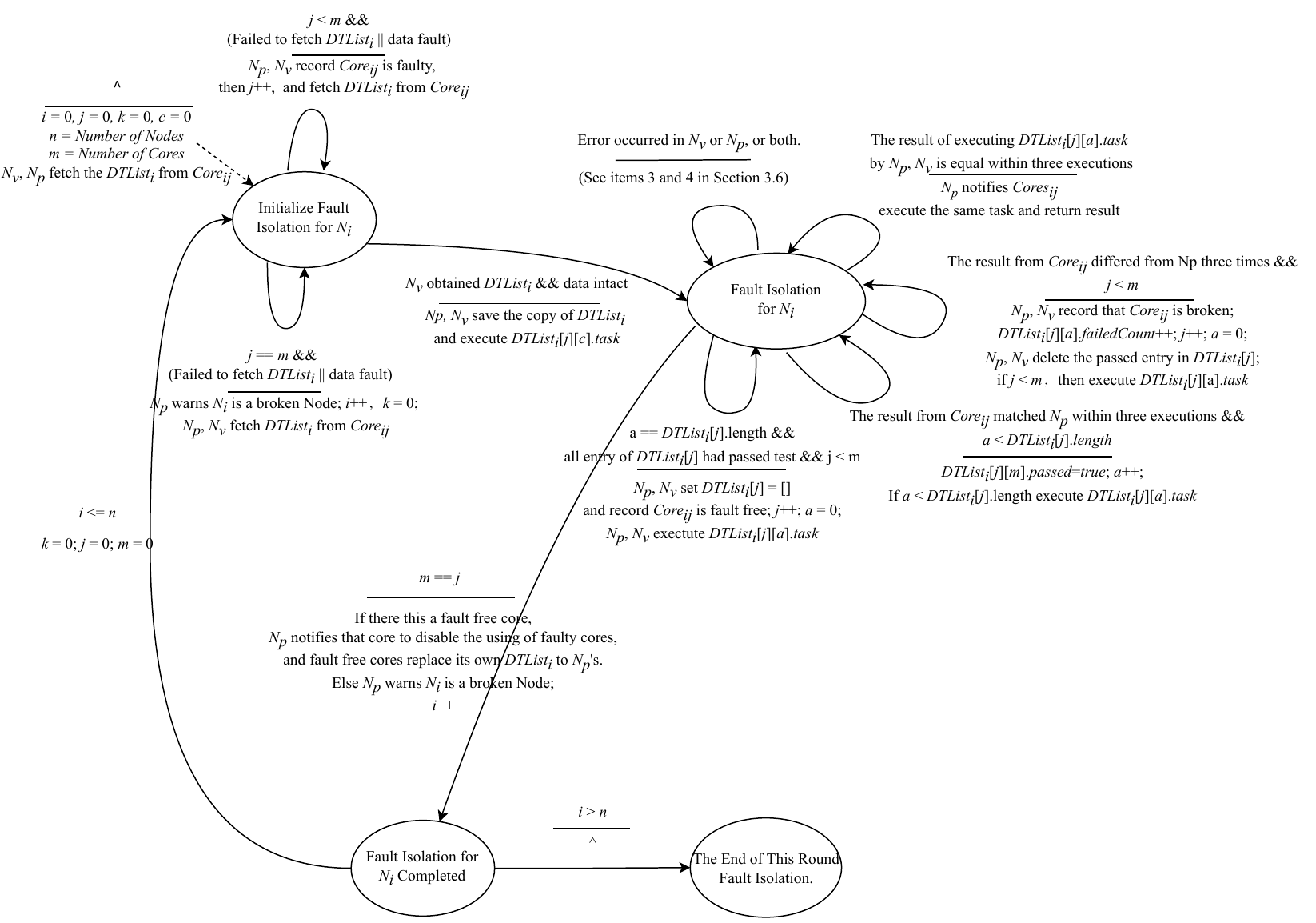}
  \caption{Finite State Machine description of the Fault Isolation process}\label{fig:fifull}
\end{figure*}

The primary and secondary leaders perform fault isolation sequentially on each core of all nodes in the system. We begin with a brief description based on the finite state machine of the fault isolation process shown in Fig.~\ref{fig:fifull}. Firstly, fault isolation for node $N_i$ is initialized. The primary and secondary leaders, $N_p$ and $N_v$, request the disputed task list $DTList_i$ from core $Core_{ik}$ (where $i$ denotes node $i$ and $k$ denotes its $k$-th core). After receiving $DTList_i$ and verifying the data integrity, $N_p$ and $N_v$ store a copy of $DTList_i$ and execute the tasks in it, comparing the results with the execution of $N_i$.

At this point, the system enters the fault isolation phase. Once fault isolation is completed, $N_p$ and $N_v$ determine the status of each core of $N_i$. The primary leader $N_p$ then instructs the fault-free cores to disable the faulty ones (the behavior of core disabling is defined in Section~\ref{subsec:tgftitmr}, Item~\ref{item:logicalIsolation}), re-enable any mistakenly disabled cores, and replace $DTList_i$ with the updated version processed by $N_p$. If all cores of $N_i$ are faulty, an alert is raised indicating that $N_i$ is completely abnormal. Once the fault isolation for $N_i$ is completed, the process proceeds to node $N_{i+1}$, and this continues until all nodes in the system have been examined. Subsequently, we describe in detail the operations performed at each stage in Figure.~\ref{fig:fifull}.


The initialization of fault isolation for node $N_i$ is illustrated in top left area of Fig 5. If $N_p$ and $N_v$ request $DTList_i$ from $Core_{ik}$ but either do not receive a response or the returned data is corrupted, $N_p$ records the core as faulty and requests $DTList_i$ from $Core_{i,k+1}$. If all cores are queried sequentially and either no response is received or the data is corrupted, $N_p$ raises an alert indicating that the entire node has failed. If $DTList_i$ is received and the data is intact, $N_p$ and $N_v$ store a copy of $DTList_i$ and execute $DTList_i[j][m].task$, which represents the $m$-th disputed task generated on the $j$-th core. The system then start the fault isolation for $N_i$.

The fault isolation phase is illustrated in top right area of Fig 5. If the results of three consecutive executions on $Core_{ij}$ are consistent with $N_p$, $N_p$ and $N_v$ execute $DTList_i[m+1]$ and remove the entry $DTList_i[m]$. If all tasks in $DTList_i$ passed test, $\text{Core}_{ij}$ is recorded as fault-free.

If the results are inconsistent, $N_p$ and $N_v$ mark $Core_{ij}$ as faulty and increment $DTList_i[m].failedCount$, which is used for exponential backoff to prevent frequent rechecking of the same core. $N_p$ and $N_v$ then execute $DTList_i[j+1][m].\text{task}$, i.e., checking $\text{Core}_{i,j+1}$, and repeat this process until all cores of $N_i$ have been examined.

If there are available cores in $N_i$, let $E_i$ denote the set of cores with permanent faults and $A_i$ the set of fault-free cores. $N_p$ replaces $DTList_i$ on the cores in $A_i$ and instructs them to disable the cores in $E_i$.

If all cores of $N_i$ are faulty, an alert is raised indicating that $N_i$ is completely abnormal. After the fault isolation for $N_i$ is completed, the system proceeds to node $N_{i+1}$. This process continues until all nodes in the system are inspected.

\subsection{The Guarantees of FTI-TMR}\label{subsec:tgftitmr}
The FTI-TMR algorithm provides highly reliable guarantees for leader election and system consistency under the Accidental-Behavior Fault Tolerance (ABFT) model. Core reliability is driven by the following mechanisms:
\begin{enumerate}
  \item{Term Mechanism: The term mechanism synchronizes the state of all cores along the logical time dimension. It effectively isolates cores whose states are corrupted due to crashes or restarts, ensuring that leader election and state synchronization occur only among honest and temporally synchronized cores, thereby enhancing the system's crash resilience.}
  \item{Majority Consensus: Since faulty cores are limited to accidental behavior rather than malicious behavior, all healthy cores honestly compute and broadcast their real $SS$ values. Assuming that more than half of the cores are operational, under the constraints of the Term mechanism and $SS$ recomputation verification, the two most stable nodes, $N_p$ and $N_v$, will obtain the majority of votes. The elected primary leader is recognized by all healthy cores and is physically among the most stable nodes, ensuring the consistency and validity of the election.}
  \item{Dynamic Self-Healing of Leaders: Even if $N_p$ or $N_v$ are initially misselected, or if one of them fails during the fault detection phase (i.e., executing a contention task three times inconsistently), $N_p$ and $N_v$ immediately trigger re-election and introduce a third highly stable node for arbitration and replacement. If no replacement is possible, the original $N_p$ and $N_v$ are removed and a new leader is re-elected (see the heartbeat mechanism).}
  \item {Heartbeat Mechanism: During permanent fault isolation, $N_p$ and $N_v$ broadcast heartbeats to the other nodes to indicate the node currently under inspection. The nodes in the system shall initiate a new election after a random delay of 150–300 ms (ensure only one node initiates the election) if any of the following conditions is met:
  \begin{itemize}
    \item {A heartbeat from $N_p$ or $N_v$ is not received within the timeout period.}
    \item {$N_p$ and $N_v$ inspect different nodes at the same time.}
    \item {The term is inconsistent.}
    \item {Any combination of the above events occurs.}
  \end{itemize}
  Once a majority of the nodes agrees, the original $N_p$ and $N_v$ are removed in next election, and the system proceeds to elect new leaders.
  }
  \item {Logical Isolation: For a node $N_i$, given the set of permanent faulty cores $E_i$ and fault-free cores $A_i$ (including $Core_{ia}$), logical isolation mitigates fault propagation. $N_p$ directs all cores in $A_i$ to adhere to schedules from $Core_{ia}$, assigning tasks only within $A_i$ and never to $E_i$. Concurrently, $Core_{ia}$ is try instruct $E_i$ from executing tasks. Consequently, any attempt by $E_i$ to assign tasks to $A_i$ is rejected, effectively isolating the accidental behavior of the faulty cores.}\label{item:logicalIsolation}
\end{enumerate}
From above, for $n$ nodes in the system, if $\lfloor\frac{n}{2}\rfloor+1$ of nodes are fault-free, the system can fail only if event in equation~\eqref{prove} happens, which is nearly impossible. 
\begin{equation}
  \label{prove}
  \begin{split} 
  & Faulty~node~becomes~leader~despite~verification~\cap \\
  & Dynamic~self-healing~of~leaders~fails~\cap \\
  & N_p~and~N_v~send~their~heartbeats~on~schedule~\cap \\
  & The~terms~of~N_p~and~N_v~match~the~system's~\cap \\
  & N_p~and~N_v~always~inspect~the~identical~node.~\cap \\
  \end{split}
\end{equation}
Therefore, for $n$ nodes in the system, FTI-TMR can always guarantee the reliable operation of the system as long as the $\lfloor\frac{n}{2}\rfloor+1$ of nodes are fault-free.

\begin{figure}[t]
  \centering
  \includegraphics[width=3.4in]{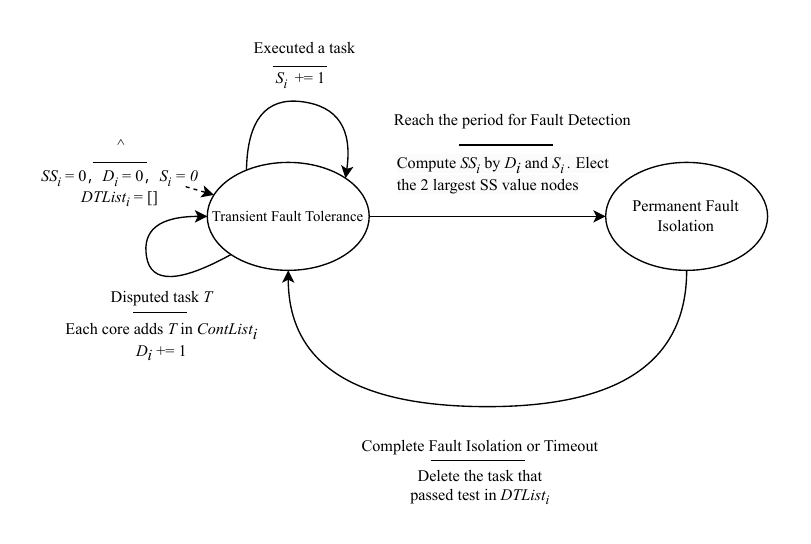}
  \caption{
    Finite State Machine description of the FTI-TMR process.
  }\label{fig:wholeProcessFDI}
\end{figure}

Fig.~\ref{fig:wholeProcessFDI} shows the overall finite state machine of the FTI-TMR process, which integrates transient fault handling, stability scoring, leader election, and permanent fault detection and isolation. The system continuously cycles through these phases to maintain high reliability and fault tolerance.

\section{Experiment}
\subsection{Experiment Setup}\label{subsec:es}
\subsubsection{Simulation Framework}\label{subsubsec:sf}
We simulated a quad-core CPU running tasks with different methods on our self-developed web-based application. A limitation of this approach is its inability to directly emulate actual hardware-level energy consumption changes and physical fault characteristics. To evaluate the effectiveness of our design in tolerating various faults, we modeled the occurrence of both transient and permanent faults within the simulation framework. For modeling transient faults, similar to previous work\cite{RN58, RN24, RN82}, we used a Poisson distribution with an average occurrence rate of $\lambda$.
\begin{equation}
  \label{eq:transientFault}
  \lambda = \lambda_0 \times 10^{\frac{d \times (1 - S)}{1 - S_{\min}}}
\end{equation}
Transient faults are triggered according to Equation~\eqref{eq:transientFault}. In prior work, $S$ represents the voltage/frequency level, selected from a set of eight values $\{\frac{0.85}{1.55}, \frac{0.95}{1.55}, \ldots, \frac{1.55}{1.55 }\}$. Due to limitations in our experimental setup, we set $S = \frac{1.45}{1.55}$ based on a comparative analysis with existing studies. Here, $\lambda_0 = 10^{-6}$ faults/second denotes the transient fault rate at the maximum voltage, and $d = 3$ is a technology-dependent constant, both adopted from established research\cite{RN58, RN24, RN82, RN2}. Accordingly, based on the properties of the Poisson distribution, the probability of a task encountering a transient fault is given by:
\begin{equation}
  \label{eachTaskTFault}
  F=1-e^{-\lambda\times T}
\end{equation}

The probability that a task executes without encountering any transient faults is: $1-F=e^{-\lambda\times T}$.

For permanent faults, a faulty core has a near-zero probability of producing a correct task result. Additionally, during critical processes such as elections and fault isolation, it has a 50\% probability of performing accidental actions.
During the experiments, we configured nine virtual 4-core machines on the simulation platform. Four of these machines contained permanently damaged cores, with the number of faulty cores being 1–4, respectively. The other five machines had no permanent core faults. 

\subsubsection{Evaluation Metrics}\label{subsubsec:emetrics}
To evaluate the reliability of the entire application system, we adopted the \textit{Probability of Failure (PoF)} \cite{RN58, RN24, RN82, RN2} as the metric, which reflects the risk level of the system producing erroneous results during execution. To ensure an Application is processed correctly, all tasks from that Application must ultimately output correct results. According to the mechanism described in Section~\ref{sec:relatedwork}, a task is considered successfully executed if at least two of its three copies encounter no failures. Conversely, \textit{if two or more copies of a task fail, the application execution is considered failed}. Therefore, \textit{PoF} is calculated as the ratio of failed applications to the total number of executed applications. Since our simulation platform cannot effectively simulate energy consumption, we used the number of executed tasks as a substitute measure for energy consumption.

\subsubsection{Applications}\label{subsubsec:app}
We simulated multiple real-world embedded applications using the standard task graph (STG)\cite{RN70}. The STG application suite covers typical scenarios such as robotic control, sparse matrix solvers, and SPEC fpppp programs. For comparison, we also included a highly parallel application consisting of 200 randomly generated discrete tasks. In each experiment, all nodes repeatedly executed the same application. For each application, we executed it 100 times on every node. Although a limited number of runs is unlikely to trigger transient faults, this setup is acceptable as our study focuses primarily on the detection and isolation of permanent faults.

\subsection{Evaluated Methods}\label{subsec:emethod}
We evaluated the following methods:
\begin{itemize}
  \item {Conventional TMR (C-TMR): In the C-TMR scheme, the system consistently executes three copies of each task, as described in Section~\ref{subsec:C-TMR}.}
  \item {Enhanced Two-Phase TMR (TP-TMR+): This method incorporates a straightforward enhancement to tolerate permanent faults at the cost of increased energy consumption. By employing an improved offline scheduling strategy, it avoids assigning multiple copies of the same task to the same core, thereby achieving fault tolerance against permanent failures}
  \item {Reactive TMR (R-TMR): Based on execution histories across different cores, this approach attempts to disable faulty cores and reassign their tasks to functional ones.}
  \item {Fault Tolerance and Isolation TMR (FTI-TMR): Our proposed fault-tolerant algorithm leverages execution history and other information to vote for the most stable devices within the interconnected system. More reliable nodes are then tasked with error detection and isolation in others, leading to improved fault tolerance and isolation capabilities. In our experiments, the initial fault isolation cycle was set to 2 rounds of application execution. After each isolation phase, the cycle length grows exponentially until it reaches a maximum of 100 rounds, at which point the growth stops. Overhead operations generated by the algorithm, such as voting and fault isolation, are also recorded as executed tasks.When the FTI-TMR mechanism reached its fault detection cycle, the values of $f_i$ and $F_0$ were set to 0, and $SS_i$ was calculated according to Equation (1).}
\end{itemize}

\begin{figure}[t]
  \centering
  \includegraphics[width=1\linewidth]{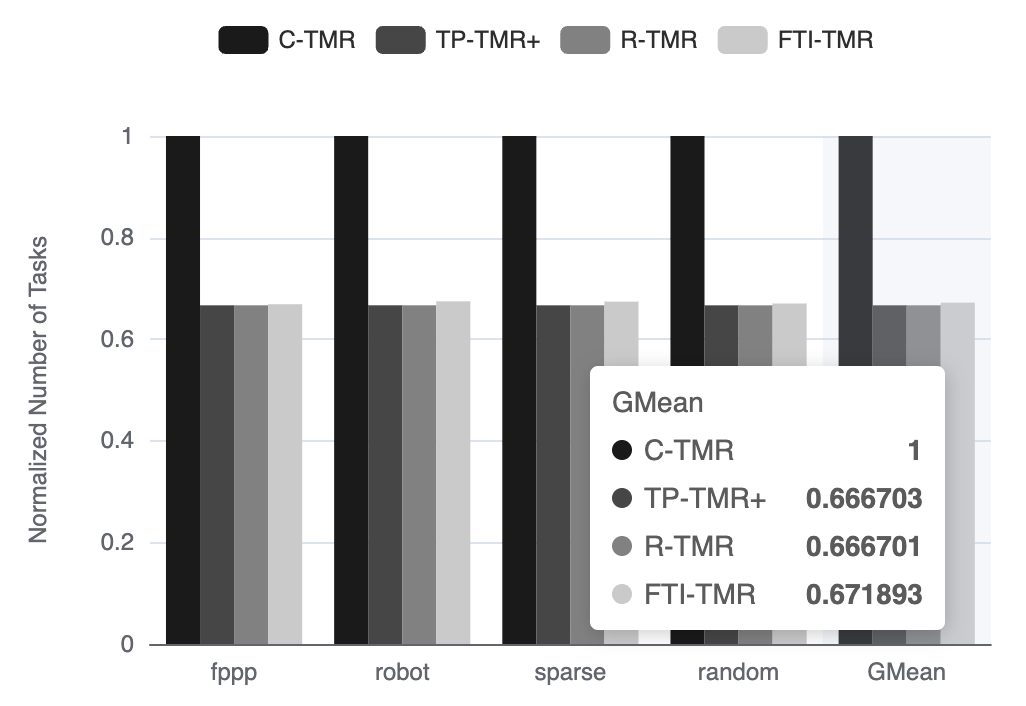}
  \caption{
    Number of executed tasks per method under no permanent faults on 9 nodes (Normalized to C-TMR).
  }\label{fig:faultfreeExperiment}
\end{figure}

\subsection{Results and Analysis}\label{subsec:ra}

Fig.~\ref{fig:faultfreeExperiment} shows the number of executed tasks for each method across nine nodes in the absence of permanent faults, normalized against the execution count of C-TMR. 
As observed, C-TMR yields the highest number of executed tasks. 
Under low transient fault rates, TP-TMR+, R-TMR, and FTI-TMR seldom execute the third task copy, thereby reducing the computational load by approximately one-third compared to C-TMR, with an average task execution count around 66\% of that of C-TMR.

FTI-TMR executes slightly more tasks than TP-TMR+ and R-TMR, as it performs additional operations for voting and fault detection at each error-checking cycle. 
Specifically, FTI-TMR's task count is about 0.76\% higher than that of the state-of-the-art R-TMR.

\begin{table}
  \centering
  \caption{Ability of each method to handle permanent core failures in m-core nodes.}
  \begin{tabular}{c m{17ex} m{13ex} m{11ex} }
  \toprule
  Method & Isolating \textless~m-1 permanent faulty cores & Isolating m-1 permanent faulty cores & Detecting m permanent faulty cores  \\
  \midrule
  C-TMR & $\times$ & $\times$ & $\times$   \\
  TP-TMR+ & $\times$ & $\times$ & $\times$   \\
  R-TMR & $\checkmark$ & $\times$ & $\times$   \\
  FTI-TMR & $\checkmark$ & $\checkmark$ & $\checkmark$  \\
  \bottomrule
  \end{tabular}\label{table:ability}
\end{table}

\begin{figure}[t]
  \centering
  \includegraphics[width=1\linewidth]{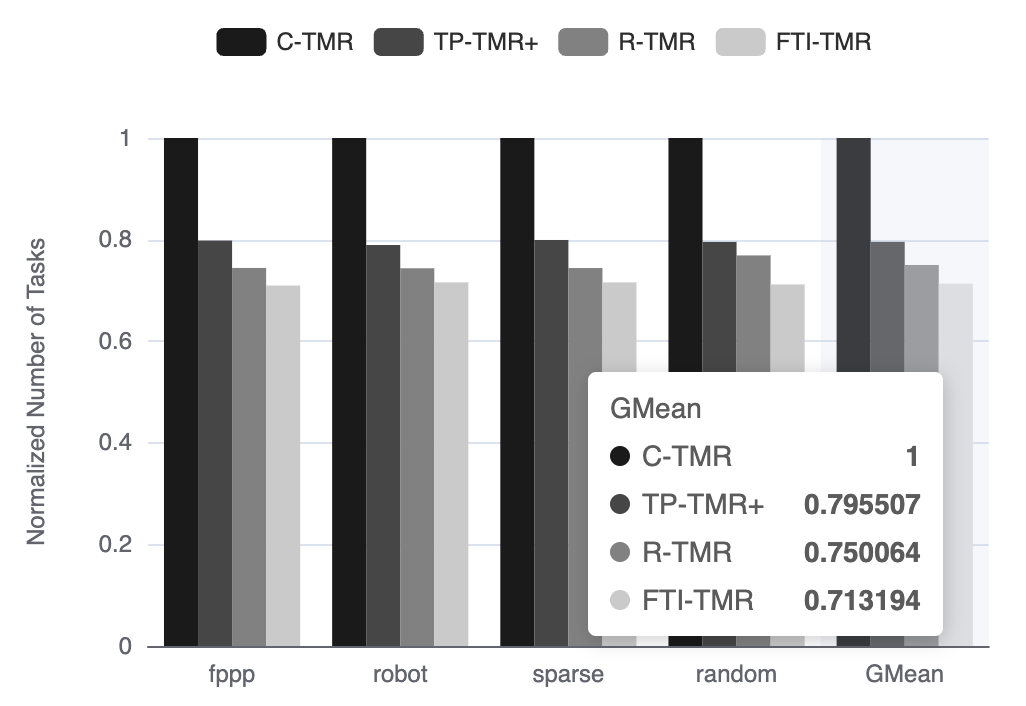}
  \caption{
    Number of executed tasks per method on 9 nodes under permanent faults (5 Fault-Free, 4 with 1–4 cores broken, normalized to C-TMR).
  }\label{fig:faultExperiment}
\end{figure}

In scenarios with permanent faults, we configured nine four-core nodes, among which four nodes had 1–4 permanently faulty cores, respectively, and five nodes remained fault-free. 
Figure~\ref{fig:faultExperiment} presents the normalized task counts relative to C-TMR under such conditions. 
All three methods---TP-TMR+, R-TMR, and FTI-TMR---show increased task counts compared to the fault-free case. 
FTI-TMR exhibits the smallest increase, followed by R-TMR and then TP-TMR+. 
As summarized in Table~\ref{table:ability}, R-TMR can detect and isolate any number of faulty cores within the system
as long as at least $2$ fault-free cores are available. This is
because, in the presence of at least two fault-free cores, R-TMR’s scheduling strategy can always allocate a task’s two copies to these fault-free cores. The third task copy might be scheduled to a faulty core. Obviously, when fewer than $2$ fault-free cores, R-TMR failes rapidly. Whereas FTI-TMR can handle up to \(m\) faulty cores, as also demonstrated in our online visual experiments. When all cores in a node become faulty, FTI-TMR can report the node as entirely faulty. Since nodes with all cores faulty require frequent execution of the third task copy, FTI-TMR's task count increases under permanent faults, yet it still reduces the number of executed tasks by about 4.92\% compared to the R-TMR.

\begin{table}
  \centering
  \caption{Probability of Failure (PoF) comparison on 9 nodes under permanent fault (5 Fault-Free, 4 with 1–4 Cores Broken).}
  \begin{tabular}{c c c c c}
  \toprule
  \multirow{2}{*}{\raisebox{-0.5ex}{Application}} & \multicolumn{4}{c}{Probability of Failure}  \\
  \cmidrule{2-5}
                        & C-TMR & TP-TMR+ & R-TMR & FDI-TMR \\
                        
  \midrule
  fppp & 0.3378 & 0.3333 & 0.2244  & 0.1211 \\
  robot & 0.3344 & 0.3356 & 0.2244  & 0.1156 \\
  sparse & 0.3344 & 0.3367 & 0.2244  & 0.1167 \\
  random & 0.3344 & 0.3333 & 0.2988  & 0.1156 \\
  \midrule
  \textbf{Average} & 0.3353 & 0.3347 & 0.2431 & 0.1172 \\
  \bottomrule
  \end{tabular}\label{table:pof}
\end{table}

We further compare the \textit{Probability of Failure} (\textit{PoF}) under this configuration, as listed in Table~\ref{table:pof}. The R-TMR scheme slightly increases the \textit{POF} under random Application workloads. This comes at the cost of delaying the allocation of two task copies to fault-free cores, thereby postponing the detection and isolation of the faulty core.
FTI-TMR achieves the lowest \textit{PoF}, benefiting from its superior fault detection and isolation capability. 
However, when a node is fully faulty, FTI-TMR cannot shut it down and only notifies about its failure, resulting in a PoF of 11.72\% in such cases. 
On average, FTI-TMR achieves a reduction of approximately 51.79\% in the \textit{PoF} compared to R-TMR.

\section{Conclusion}
As a widely adopted fault-tolerance mechanism, Triple Modular Redundancy (TMR) ensures high reliability through triple redundant computation and majority voting. However, traditional TMR incurs significant energy overhead, making it challenging to deploy directly in energy-constrained systems such as embedded or avionics applications. To mitigate this, various improved schemes have been proposed—including task scheduling and power management optimizations—among which Two-Phase TMR (TP-TMR) and Reactive TMR (R-TMR) are notable examples.

TP-TMR introduces a energy-optimized triple modular redundancy method that executes the third task copy on demand. Building upon this, R-TMR allocates tasks to different cores (the Enhanced Two-Phase TMR) and incorporates an additional lightweight hardware to identify and isolate permanent faults, cause additional hardware overhead. And R-TMR is difficulties in detecting and isolting multiple simultaneous permanent faults.~They both remain confined to the isolated node paradigm—relying on local hardware.

To overcome these limitations, this paper proposes Fault Tolerance and Isolation TMR (FDI-TMR), a method designed for interconnected systems. Under normal operation, FDI-TMR employs the Enhanced Two-Phase TMR (TP-TMR+) mechanism to save energy and tolerate faults, thereby improving energy-efficiency. To detect and isolate permanent faults, the method introduces a stability score that assesses the stable status of each node, allowing more reliable nodes to perform periodic permanent fault detection and isolation on others. This approach achieves superior permanent fault coverage, significantly improving both reliability and energy efficiency.

A key limitation of this paper is: our work was done on a web simulation platform. It can't capture the physics of hardware-level energy drain or how permanent faults really behave. Nevertheless, we think the core idea behind FDI-TMR—having reliable nodes actively perform diagnosis on others—is not limited to CPU fault tolerance based on TMR. With the rapid advancement of artificial intelligence and autonomous systems, we're excited to try applying this same ``reliable-node'' concept to multi-robot teams. The idea would be to have more dependable robots monitor, diagnose, and maybe even repair other units in the field. This could be a step toward building truly autonomous systems.

%

\bibliographystyle{IEEEtran}
\bibliography{IEEEabrv,refs}

\end{document}